\documentclass[12pt]{article}
\usepackage{epsfig}
\begin{document}
\title{Study of in-medium $\eta^\prime$ properties in the ($\gamma$, ${\eta^\prime}p$)
 reaction\\
 on nuclei}
\author{E. Ya. Paryev$^{1,2}$\\
{\it $^1$Institute for Nuclear Research, Russian Academy of Sciences,}\\
{\it Moscow 117312, Russia}\\
{\it $^2$Institute for Theoretical and Experimental Physics,}\\
{\it Moscow 117218, Russia}}

\renewcommand{\today}{}
\maketitle

\begin{abstract}
We study the near-threshold photoproduction of $\eta^\prime$ mesons from nuclei
in coincidence with forward going protons in the kinematical conditions of the Crystal Barrel/TAPS
experiment, recently performed at ELSA. The calculations have been performed
within a collision model based on the nuclear spectral function.
The model accounts for both the primary ${\gamma}p \to {\eta^\prime}p$ process and the
two-step intermediate nucleon rescattering processes as well as the effect of the nuclear $\eta^\prime$
mean-field potential. We calculate the exclusive $\eta^\prime$ kinetic energy distributions for the
$^{12}$C($\gamma$, ${\eta^\prime}p$) reaction for different scenarios of $\eta^\prime$ in-medium
modification. We find that the considered two-step rescattering mechanism plays an insignificant role in
${\eta^\prime}p$ photoproduction off the carbon target. We also demonstrate that the calculated $\eta^\prime$
kinetic energy distributions in primary photon--proton ${\eta^\prime}p$ production reveal strong
sensitivity to the depth of the real ${\eta^\prime}$ potential at normal nuclear matter density (or to the
${\eta^\prime}$ in-medium mass shift) in the studied incident photon energy regime. Therefore, such
observables may be useful to help determine the above ${\eta^\prime}$ in-medium renormalization from the comparison
of the results of our calculations with the data from the CBELSA/TAPS experiment. In addition, we show that
these distributions are also strongly influenced by the momentum-dependent optical potential, which the outgoing
participant proton feels inside the carbon nucleus. This potential should be taken into account
in the analysis of these data with the aim to obtain information on the ${\eta^\prime}$ modification in cold
nuclear matter.
\end{abstract}

\newpage

\section*{1. Introduction}

\hspace{1.5cm} The investigation of the ${\eta^\prime}$ meson mass in nuclear matter and the strength of
its inelastic interaction with nucleons has received considerable interest in recent years
(see, for example, [1--21]) in the context of obtaining valuable information both on the partial
restoration of chiral symmetry at a finite density and on the behavior of the $U_A(1)$ anomaly in the
nuclear medium as well as in view of the possible existence of such exotic nuclear systems as
${\eta^\prime}$-mesic nuclei. Recent inclusive ${\eta^\prime}$ photoproduction measurements of the
CBELSA/TAPS Collaboration show [20, 21] that the ${\eta^\prime}$-nucleus optical potential
$V_{opt}=V_{real}+iW$ is $V_{real}(\rho_0)=-(37{\pm}10(stat.){\pm}10(syst.))$ MeV, which is equal to the
${\eta^\prime}$ meson mass shift in the medium, and $W(\rho_0)=-(10{\pm}2.5)$ MeV at the saturation density
$\rho_0$ and for average ${\eta^\prime}$ momenta of $\approx$ 1 GeV/c.
The ${\eta^\prime}$ in-medium mass shift, determined by the CBELSA/TAPS Collaboration, is very
similar to that of -37 MeV, calculated within the Quark Meson Coupling model [7--9], and disfavors the larger
${\eta^\prime}$ mass shifts, downwards by up to 80--200 MeV, predicted in [2, 3, 5, 12--15]. The small imaginary
part of the ${\eta^\prime}$-nucleus optical potential compared to its real part provides the possibility for the
observation of relatively narrow bound ${\eta^\prime}$-nucleus states [2--4, 6--8, 10--12]. The search of
such states by use of the ($p$, $d$) reaction on $^{12}$C has been very recently undertaken at FRS/GSI and is
planned at FAIR [22] as well as in photoproduction at ELSA [23].

  In an attempt to deduce the real part of the ${\eta^\prime}$-nucleus potential at lower ${\eta^\prime}$
momenta  the near-threshold exclusive photoproduction of ${\eta^\prime}$ mesons from a carbon target nucleus
has been studied in coincidence with forward going protons by the CBELSA/TAPS Collaboration [24].
In this respect, the main purpose
of the present work is to get estimates of the absolute ${\eta^\prime}p$ yield from the
$^{12}$C($\gamma$, ${\eta^\prime}p$) reaction in the kinematical conditions of the CBELSA/TAPS experiment [24].
The calculations are based on a collision model [19] developed for the analysis of the inclusive data [20, 21]
on the transparency
ratio for ${\eta^\prime}$ mesons as well as on their momentum distribution and excitation function in
${\gamma}A$ collisions and expanded to take the additional production of protons in these collisions into account
in different scenarios for the ${\eta^\prime}$ meson in-medium mass shift. In view of the expected data from this
experiment, the estimates can be used as an important tool for determining the above shift in cold nuclear matter.

\section*{2. The formalism}

\section*{2.1. Direct  ${\eta^\prime}p$ production mechanism}

\hspace{1.5cm} Following [19], we assume that only the elementary process
\begin{equation}
\gamma+p \to \eta^\prime+p
\end{equation}
with the lowest free production threshold ($\approx$ 1.446 GeV) contributes to the direct
production of the ${\eta^\prime}p$ pairs on a carbon target nucleus in the bombarding energy
range of our interest. As in [19], we will employ in our calculations of pair production
from the primary process (1) for the in-medium mass of the ${\eta^\prime}$ mesons
their average in-medium mass $<m^*_{{\eta^\prime}}>$ defined in view of equations (3), (4) from
[19] as:
\begin{equation}
<m^*_{{\eta^\prime}}>=m_{{\eta^\prime}}+V_0\frac{<{\rho_N}>}{{\rho_0}},
\end{equation}
where $m_{{\eta^\prime}}$ is the ${\eta^\prime}$ free space mass, $<{\rho_N}>$ is the
average nucleon density ($<{\rho_N}>={\rho_0}/2$) and $V_0$ is the ${\eta^\prime}$ scalar
potential depth (or the ${\eta^\prime}$ in-medium mass shift) at saturation density $\rho_0$.
For the quantity $V_0$ we will employ the seven options: i) $V_0=0$, ii) $V_0=-25$ MeV,
iii) $V_0=-50$ MeV, iv) $V_0=-75$ MeV, v) $V_0=-100$ MeV, vi) $V_0=-125$ MeV, and vii) $V_0=-150$ MeV
throughout the following study.
The total energy $E^\prime_{{\eta^\prime}}$ of the ${\eta^\prime}$ meson inside the nuclear medium is
expressed via its average effective mass $<m^*_{{\eta^\prime}}>$ and its in-medium momentum
${\bf p}^{\prime}_{{\eta^\prime}}$ by means of equation:
\begin{equation}
E^\prime_{{\eta^\prime}}=\sqrt{({\bf p}^{\prime}_{{\eta^\prime}})^2+(<m^*_{{\eta^\prime}}>)^2}.
\end{equation}
This energy is equal to the total vacuum energy $E_{{\eta^\prime}}$ of the ${\eta^\prime}$ meson
having the vacuum momentum ${\bf p}_{{\eta^\prime}}$:
\begin{equation}
E^\prime_{{\eta^\prime}}=E_{{\eta^\prime}}=
\sqrt{{\bf p}^2_{{\eta^\prime}}+m^2_{{\eta^\prime}}}.
\end{equation}
We also take into account in these calculations the
medium modification of the final proton by using, by analogy with equation (2), its average in-medium mass
$<m^*_{N}>$:
\begin{equation}
<m^*_{N}({\bf p}_N^{'2})>=m_{N}+V_{NA}^{\rm SC}({\bf p}_N^{'2})\frac{<{\rho_N}>}{{\rho_0}}.
\end{equation}
Here, $m_{N}$ is the nucleon free space mass and $V_{NA}^{\rm SC}({\bf p}_N^{'2})$ is the scalar momentum-dependent
nuclear nucleon potential at saturation density. It depends on the in-medium nucleon momentum ${\bf p}_N^{'}$.
The potential $V_{NA}^{\rm SC}({\bf p}_N^{'2})$ can be determined from the relation
\begin{equation}
\sqrt{m_N^2+{\bf p}_N^{'2}}+V_{NA}^{\rm SEP}({\bf p}_N^{'2})\frac{{\rho_N}({\bf r})}{{\rho_0}}=
\sqrt{\left[m_N+V_{NA}^{\rm SC}({\bf p}_N^{'2})\frac{{\rho_N}({\bf r})}{{\rho_0}}\right]^2+{\bf p}_N^{'2}},
\end{equation}
where $V_{NA}^{\rm SEP}({\bf p}_N^{'2})$ is the Schroedinger equivalent potential for nucleons. This potential is
shown in figure 1 of [25] at density $\rho_0$ as a function of the momentum relative to the nuclear matter at
rest. It can be parametrized as follows:
\begin{equation}
V_{NA}^{\rm SEP}({\bf p}_N^{'2})=\left(V_1-V_2{\rm e}^{-2.3{\bf p}_N^{'2}}\right);\\\\ V_1=50~{\rm MeV},
\\\\V_2=120~{\rm MeV},
\end{equation}
where the momentum $|{\bf p}_N^{'}|$ is measured in GeV/c. Taking the square of equation (6) and neglecting
the terms proportional to the $\left(\frac{{\rho_N}({\bf r})}{{\rho_0}}\right)^2$ (this introduces an error of the order of 1--2 \%), we obtain that
\begin{equation}
V_{NA}^{\rm SC}({\bf p}_N^{'2})=\frac{\sqrt{m_N^2+{{\bf p}_N^{'2}}}}{m_N}V_{NA}^{\rm SEP}({\bf p}_N^{'2}).
\end{equation}
The total energy $E_N^{'}$ of the outgoing proton in the nuclear interior can be expressed in terms of its
effective mass $<m^*_{N}({\bf p}_N^{'2})>$ defined above and its in-medium momentum ${\bf p}_N^{'}$ as
in the free particle case, namely:
\begin{equation}
E_{N}^{'}=\sqrt{[<m^*_{N}({\bf p}_{N}^{'2})>]^2+{\bf p}_{N}^{'2}}.
\end{equation}
When the proton escapes from the nucleus with momentum ${\bf p}_{N}$ its total energy $E_N$ becomes
equal to that corresponding to its bare mass $m_N$: $E_{N}=\sqrt{m_{N}^2+{\bf p}_{N}^{2}}$.
As in the ${\eta^\prime}$ meson case, it is natural to assume that $E_N^{'}=E_N$.

  Neglecting the distortion of the incident photon in nuclear matter
and describing the ${\eta^\prime}$ meson and proton final-state absorption by the effective cross sections
$\sigma_{{\eta^\prime}N}$ and $\sigma_{pN}^{\rm tot}$ as well as using the
results given in [19], we can represent the exclusive differential
cross section for the production of ${\eta^\prime}$ meson with the vacuum momentum ${\bf p}_{{\eta^\prime}}$
in coincidence with the proton with the vacuum momentum ${\bf p}_{N}$
off nuclei in the primary photon--induced reaction channel (1) as follows:
\begin{equation}
\frac{d\sigma_{{\gamma}A\to {\eta^\prime}pX}^{({\rm prim})}
(E_{\gamma})}
{d{\bf p}_{\eta^\prime}d{\bf p}_{N}}=I_{V}[A,\theta_{\eta^\prime}]
\left(\frac{Z}{A}\right)\left<\frac{d\sigma_{{\gamma}p\to {\eta^\prime}p}({\bf p}_{\gamma};
{\bf p}^{\prime}_{\eta^\prime},{\bf p}_N^{'})}{d{\bf p}^{\prime}_{\eta^\prime}d{\bf p}_N^{'}}\right>_A
\frac{d{\bf p}^{\prime}_{{\eta^\prime}}}{d{\bf p}_{{\eta^\prime}}}
\frac{d{\bf p}_{N}^{'}}{d{\bf p}_{N}},
\end{equation}
where
\begin{equation}
I_{V}[A,\theta_{\eta^\prime}]=A\int\limits_{0}^{R}r_{\bot}dr_{\bot}
\int\limits_{-\sqrt{R^2-r_{\bot}^2}}^{\sqrt{R^2-r_{\bot}^2}}dz
\rho(\sqrt{r_{\bot}^2+z^2})
\exp{\left[-\sigma_{pN}^{\rm tot}A\int\limits_{z}^{\sqrt{R^2-r_{\bot}^2}}
\rho(\sqrt{r_{\bot}^2+x^2})dx\right]}
\end{equation}
$$
\times
\int\limits_{0}^{2\pi}d{\phi}\exp{\left[-\sigma_{{\eta^\prime}N}A\int\limits_{0}^{l(\theta_{\eta^\prime},\phi)}
\rho(\sqrt{x^2+2a(\theta_{\eta^\prime},\phi)x+b+R^2})dx\right]},
$$
\begin{equation}
a(\theta_{\eta^\prime},\phi)=z\cos{\theta_{\eta^\prime}}+r_{\bot}\sin{\theta_{\eta^\prime}}\cos{\phi},\\\
b=r_{\bot}^2+z^2-R^2,
\end{equation}
$$
l(\theta_{\eta^\prime},\phi)=\sqrt{a^2(\theta_{\eta^\prime},\phi)-b}-
a(\theta_{\eta^\prime},\phi)
$$
and
\begin{equation}
\left<\frac{d\sigma_{{\gamma}p\to {\eta^\prime}p}({\bf p}_{\gamma};
{\bf p}_{\eta^\prime}^{'},{\bf p}_N^{'})}
{d{\bf p}_{\eta^\prime}^{'}d{\bf p}_N^{'}}\right>_A=
\int\int
P_A(|{\bf p}_t|,E)d{\bf p}_tdE
\end{equation}
$$
\times
\left\{\frac{d\sigma_{{\gamma}p\to {\eta^\prime}p}[\sqrt{s},<m^*_{{\eta^\prime}}>,
<m^*_{N}({\bf p}_{N}^{'2})>,{\bf p}_{\eta^\prime}^{'},{\bf p}_N^{'}]}
{d{\bf p}_{\eta^\prime}^{'}d{\bf p}_N^{'}}\right\}.
$$
Here,
$d\sigma_{{\gamma}p\to {\eta^\prime}p}[\sqrt{s},<m^*_{{\eta^\prime}}>,
<m^*_{N}({\bf p}_{N}^{'2})>,{\bf p}_{\eta^\prime}^{'},{\bf p}_N^{'}] /d{\bf p}_{\eta^\prime}^{'}d{\bf p}_N^{'}$
is the off-shell differential cross section for the production of ${\eta^\prime}$ meson and proton with the
in-medium momenta ${\bf p}_{{\eta^\prime}}^{'}$ and ${\bf p}_N^{'}$, respectively, in reaction (1);
$\theta_{\eta^\prime}$ is the polar angle of vacuum momentum ${\bf p}_{{\eta^\prime}}$ in the laboratory
system with the $z$ axis directed along the momentum ${\bf p}_{\gamma}$ of the initial photon,
and the other quantities, appearing in the equations (10)--(13), are defined in [19].
In equations (11) and (12) it is assumed that the paths of the produced ${\eta^\prime}$ meson and proton out
of the nucleus are not disturbed by the ${\eta^\prime}A$ and $pA$ optical potentials as well as by the
${\eta^\prime}N$ and $pN$ quasielastic rescatterings and that the final proton is going in forward direction.
In addition, the correction of the expression (13) for Pauli blocking leading to the suppression of the phase
space available for the final-state proton is disregarded. Such approximations are allowed in calculating the
${\eta^\prime}p$ production cross section for kinematics of the analyzed experiment. Thus, for instance, at
a beam energy of 2.0 GeV the proton laboratory kinetic energies (momenta) in the process
${\gamma}p \to {\eta^\prime}p$, taking place on a free proton at rest, for its laboratory production angle of $<\theta_N>=7.5^{\circ}$
\footnote{$^)$Which is the average angle of the angular range of 1$^{\circ}$--11$^{\circ}$ covered by the
Mini-TAPS forward array.}$^)$
are 48 (303) and 948 MeV (1637 MeV/c). They correspond to two cases with ${\eta^\prime}$ going forward and
proton going backward in the cm system and vice versa
\footnote{$^)$It is also worth noting that the ${\eta^\prime}$ laboratory kinetic energies (momenta) corresponding
to these two cases are 994 (1701) and 94 MeV (434 MeV/c), which indicates that the low-momentum ${\eta^\prime}$
mesons are produced together with the high-momentum protons in the chosen kinematics.}$^)$
.
The above energies are substantially greater than the average Fermi energy of
${\bar E}_F={\bar p}_F^2/2m_N{\simeq}24$ MeV of the carbon target nucleus, corresponding to the average Fermi
momentum of ${\bar p}_F=210$ MeV/c [26]. Hence, the influence of the Pauli blocking is expected to be negligible
\footnote{$^)$We have checked that it has indeed no significant influence on our results.}$^)$
.

    Accounting for the formula (16) from [19] and equation (9), we get the following expression for the
elementary in-medium differential cross section
$d\sigma_{{\gamma}p\to {\eta^\prime}p}[\sqrt{s},<m^*_{{\eta^\prime}}>,
<m^*_{N}({\bf p}_{N}^{'2})>,{\bf p}_{\eta^\prime}^{'},{\bf p}_N^{'}] /d{\bf p}_{\eta^\prime}^{'}d{\bf p}_N^{'}$:
\begin{equation}
\frac{d\sigma_{{\gamma}p\rightarrow {\eta^\prime}p}[\sqrt{s},<m^*_{{\eta^\prime}}>,<m^*_{N}({\bf p}_{N}^{'2})>,
{\bf p}_{\eta^\prime}^{'},{\bf p}_N^{'}]}
{d{\bf p}_{\eta^\prime}^{'}d{\bf p}_N^{'}}
={\frac{{\pi}}{I_2[s,<m^*_{N}({\bf p}_{N}^{'2})>,<m^*_{{\eta^\prime}}>]E^{\prime}_{{\eta^\prime}}}}
\end{equation}
$$
\times
{\frac{d\sigma_{{\gamma}p\rightarrow {\eta^\prime}p}[{\sqrt{s}},<m^*_{{\eta^\prime}}>,
<m^*_{N}({\bf p}_{N}^{'2})>,{\theta_{\eta^\prime}^*}]}
{d{\bf \Omega}_{\eta^\prime}^{*}}}
$$
$$
\times
{\frac{1}{(\omega+E_t)}}\delta\left[\omega+E_t-\sqrt{[<m^*_{N}({\bf p}_{N}^{'2})>]^2+{\bf p}_N^{'2}}\right]
\delta({\bf Q}+{\bf p}_t-{\bf p}_N^{'}),
$$
where
\begin{equation}
I_2[s,<m^*_{N}({\bf p}_{N}^{'2})>,<m^*_{{\eta^\prime}}>]=\frac{\pi}{2}
\frac{\lambda(s,<m^*_{N}({\bf p}_{N}^{'2})>^{2},<m^*_{{\eta^\prime}}>^{2})}{s},
\end{equation}
\begin{equation}
\lambda(x,y,z)=\sqrt{{\left[x-({\sqrt{y}}+{\sqrt{z}})^2\right]}{\left[x-
({\sqrt{y}}-{\sqrt{z}})^2\right]}},
\end{equation}
\begin{equation}
\omega=E_{\gamma}-E^{\prime}_{\eta^\prime}, \,\,\,\,{\bf Q}={\bf p}_{\gamma}-{\bf p}^{\prime}_{\eta^\prime}.
\end{equation}
For the
differential cross section $d\sigma_{{\gamma}p\rightarrow {\eta^\prime}p}/d{\bf \Omega}_{\eta^\prime}^{*}$
of the reaction (1) in the ${\gamma}p$ c.m.s. we use the fit (20) from [19], in which in our case
the threshold photon energy $E_{\gamma}^{\rm thr}$ should be expressed via the effective masses
$<m^*_{{\eta^\prime}}>$ and $<m^*_{N}({\bf p}_{N}^{'2})>$ as follows:
\begin{equation}
E^{\rm thr}_{\gamma}[<m^*_{{\eta^\prime}}>,<m^*_{N}({\bf p}_{N}^{'2})>]=
\frac{[<m^*_{{\eta^\prime}}>+<m^*_{N}({\bf p}_{N}^{'2})>]^2-m_N^2}{2m_N}.
\end{equation}
The $\eta^\prime$ meson production angles ${\theta_{\eta^\prime}^*}$ and ${\theta_{\eta^\prime}}$
in the ${\gamma}N$ c.m.s. and in the l.s. frame are connected through the relation (34) from [19].

  Let us now modify the expression (10), describing the respective exclusive differential cross section
for the ${\eta^\prime}p$ production in ${\gamma}A$ collisions, to that corresponding to the kinematical
conditions of the CBELSA/TAPS experiment. In this experiment, the differential cross section for production
of $\eta^\prime$ mesons in the interaction of photons of energies of 1.5--2.2 GeV,
1.5--2.6 GeV, and 1.3--2.6 GeV with $^{12}$C target
nucleus in coincidence with protons, which were required to have vacuum kinetic energies larger than
$E_{\rm kin}^{\rm N,G}=50$ MeV and to be in the polar angular range of 1$^{\circ}$--11$^{\circ}$ in l.s.,
was measured as a function of their vacuum kinetic energy $E_{\rm kin}^{\eta^\prime}$ [24].
Substituting equations (11)--(14) in (10) and performing an averaging over the incident photon energy
$E_{\gamma}$ with $1/E_{\gamma}$ weighting in the range of
${\Delta}E_{\gamma}=E_{\gamma}^{(1)}$--$E_{\gamma}^{(2)}$
($E_{\gamma}^{(1)}$--$E_{\gamma}^{(2)}=$1.5--2.2 GeV, or 1.5--2.6 GeV, or 1.3--2.6 GeV)
as well as the
integration over the full ${\eta^\prime}$ solid angle ${\bf \Omega}_{\eta^\prime}$ with accounting for that
$d{\bf p}_{\eta^\prime}=p_{\eta^\prime}E_{\eta^\prime}dE_{\rm kin}^{\eta^\prime}d{\bf \Omega}_{\eta^\prime}$,
$d{\bf p}_{\eta^\prime}^{'}/d{\bf p}_{\eta^\prime}=p_{\eta^\prime}^{'}/p_{\eta^\prime}$ [19], over the allowed
values of the final proton three-momentum ${\bf p}_N=p_N{\bf \Omega}_N$ assuming that its polar $\theta_N$
and azimuthal $\phi_N$ angles, in line with above mentioned, do not deviate from those $\theta_N^{'}$ and
$\phi_N^{'}$ of its in-medium momentum ${\bf p}_N^{'}$ and
over the angle $\theta_t$ between the momentum of the struck target proton ${\bf p}_t$
and the momentum transfer ${\bf Q}$, we can represent the above differential cross section in the following
form:
\begin{equation}
\left<\frac{d\sigma_{{\gamma}A\to {\eta^\prime}pX}^{({\rm prim})}}
{dE_{\rm kin}^{\eta^\prime}d{\bf \Omega}_{N}}\right>_{{\Delta}E_{\gamma},{\Delta}{\bf \Omega}_N}=
\frac{1}{\ln\frac{E_{\gamma}^{(2)}}{E_{\gamma}^{(1)}}}
\int\limits_{E_{\gamma}^{(1)}}^{E_{\gamma}^{(2)}}\frac{dE_{\gamma}}{E_{\gamma}}
\left<\frac{d\sigma_{{\gamma}A\to {\eta^\prime}pX}^{({\rm prim})}(E_{\gamma})}
{dE_{\rm kin}^{\eta^\prime}d{\bf \Omega}_{N}}\right>_{{\Delta}{\bf \Omega}_N},
\end{equation}
where
\begin{equation}
\left<\frac{d\sigma_{{\gamma}A\to {\eta^\prime}pX}^{({\rm prim})}(E_{\gamma})}
{dE_{\rm kin}^{\eta^\prime}d{\bf \Omega}_{N}}\right>_{{\Delta}{\bf \Omega}_N}=
\left(\frac{2\pi}{{\Delta}{\bf \Omega}_N}\right)\left(\frac{Z}{A}\right)
\int\limits_{-1}^{1}d\cos{{\theta_{\eta^\prime}}}I_{V}[A,\theta_{\eta^\prime}]
\end{equation}
$$
\times
\left(\frac{{\pi}p_{\eta^\prime}^{'}}{Q}\right)\int\limits_{0}^{p_t^{\rm max}}
\int\limits_{0}^{2\pi}\int\limits_{E_{\rm min}}^{E_{\rm max}}P_A(p_t,E)p_tdp_td\phi_tdE
$$
$$
\times
\frac{\theta(1-|x_0|)\theta[(\omega+E_t)-(E_{\rm kin}^{\rm N,G}+m_N)]}
{I_2[s(x_0,p_t,\phi_t,E),<m^*_{N}(Q^2+p_t^2+2Qp_tx_0)>,<m^*_{{\eta^\prime}}>]
F_{\rm DIF}^{({\rm prim})}(x_0,p_t,Q)}
$$
$$
\times
{\frac{d\sigma_{{\gamma}p\rightarrow {\eta^\prime}p}[{\sqrt{s(x_0,p_t,\phi_t,E)}},<m^*_{{\eta^\prime}}>,
<m^*_{N}(Q^2+p_t^2+2Qp_tx_0)>,{\theta_{\eta^\prime}^*}]}
{d{\bf \Omega}_{\eta^\prime}^{*}}}
$$
$$
\times
\theta(\cos1^{\circ}-\cos{\theta}_{{\bf Q}+{\bf p}_t})
\theta(\cos{\theta}_{{\bf Q}+{\bf p}_t}-\cos11^{\circ})
$$
and
\begin{equation}
F_{\rm DIF}^{({\rm prim})}(x_0,p_t,Q)=|1+2<m^*_{N}(Q^2+p_t^2+2Qp_tx_0)>
[\frac{V_{NA}^{\rm SEP}(Q^2+p_t^2+2Qp_tx_0)}{2m_N\sqrt{m_N^2+Q^2+p_t^2+2Qp_tx_0}}
\end{equation}
$$
+
 \frac{\sqrt{m_N^2+Q^2+p_t^2+2Qp_tx_0}}{m_N}
 \left(\frac{V_2{\cdot}2.3}{\rm GeV^2}\right){\rm e}^{-2.3(Q^2+p_t^2+2Qp_tx_0)}]
 \frac{<{\rho_N}>}{{\rho_0}}|;
$$
\begin{equation}
{\Delta}{\bf \Omega}_N=2\pi(\cos{1}^{\circ}-\cos{11}^{\circ}), \\\,Q=|{\bf Q}|=\sqrt{p_{\gamma}^2+
p_{\eta^\prime}^{'2}-2p_{\gamma}p_{\eta^\prime}^{'}\cos{\theta_{\eta^\prime}}}, \\\,\theta(x)=\frac{x+|x|}{2|x|},
\end{equation}
\begin{equation}
\cos{\theta_{{\bf Q}+{\bf p}_t}}=\frac{({\bf Q}+{\bf p}_t){\bf p}_{\gamma}}
{|{\bf Q}+{\bf p}_t||{\bf p}_{\gamma}|}=
\frac{p_{\gamma}-p_{\eta^\prime}^{'}\cos{\theta_{\eta^\prime}}+p_t\cos{\theta_x}}
{\sqrt{Q^2+p_t^2+2Qp_tx_0}},
\end{equation}
\begin{equation}
\cos{\theta_x}=\frac{{\bf p}_t{\bf p}_{\gamma}}{|{\bf p}_t||{\bf p}_{\gamma}|}=
x_0\cos{\theta_{\bf Q}}+\sqrt{1-x_0^2}\sin{\theta_{\bf Q}}\cos{(\phi_t-\phi_{\gamma})},
\end{equation}
\begin{equation}
\cos{\theta_{\bf Q}}=\frac{{\bf Q}{\bf p}_{\gamma}}{|{\bf Q}||{\bf p}_{\gamma}|}=
\frac{p_{\gamma}-p_{\eta^\prime}^{'}\cos{\theta_{\eta^\prime}}}{Q},\\\\\
E_{\rm min}=M_{A-1}+m_N-M_A,
\end{equation}
\begin{equation}
  s(x_0,p_t,\phi_t,E)=(E_{\gamma}+E_t)^2-({\bf p}_{\gamma}+{\bf p}_t)^2=
  (E_{\gamma}+E_t)^2-p_{\gamma}^2-p_t^2-2p_{\gamma}p_t\cos{\theta_x}.
\end{equation}
Here, $\phi_t$ and $\phi_{\gamma}$ are the azimuthal angles of momenta ${\bf p}_t$
and ${\bf p}_{\gamma}$ in the reference frame in which $z^{'}$ axis is directed along the
momentum transfer ${\bf Q}$ and $x_0$ is the root of an equation $f(x_0)=0$, where
\begin{equation}
f(x)=\omega+E_t-\sqrt{[<m^*_{N}(Q^2+p_t^2+2Qp_tx)>]^2+Q^2+p_t^2+2Qp_tx}, \,\,\,\, x=\cos{\theta_t}.
\end{equation}
This root has been found by using the respective numerical procedure.
In our calculations of the ${\eta^\prime}p$ differential cross section (19) on a $^{12}$C target nucleus
we have employed the values of $p_t^{\rm max}=1.0$ GeV/c and $E_{\rm max}=0.3$ GeV. They determine entirely
the region of $p_t$ and $E$, which gives the main contribution to the cross section (20).
Also, in these calculations we used the values $\sigma_{{\eta^\prime}N}=11$ mb [21] and $\sigma_{pN}^{\rm tot}=40$ mb for the relevant outgoing ${\eta^\prime}$ and proton energies, respectively.
The nuclear spectral function $P_A(|{\bf p}_t|,E)$ for $^{12}$C was taken from [27].

    Before going further, let us get the expression for the differential cross section
$\left<\frac{d\sigma_{{\gamma}p\to {\eta^\prime}p}(E_{\gamma})}
{dE_{\rm kin}^{\eta^\prime}d{\bf \Omega}_{N}}\right>_{{\Delta}{\bf \Omega}_N}$ of the reaction
${\gamma}p \to {\eta^\prime}p$ occuring on a free target proton being at rest.
It was used in our calculations of the ${\eta^\prime}p$ pair production in ${\gamma}p$ collisions.
This expression can be obtained from the more general one (20) in the limits: $A \to 1$, $Z \to 1$,
$\sigma_{{\eta^\prime}N} \to 0$, $\sigma_{pN}^{\rm tot} \to 0$, $p_t \to 0$, $E_t \to m_N$
as well as $\eta^\prime$ and $p$ in-medium mass shifts $\to 0$. As a result, after some algebra we find:
\begin{equation}
\left<\frac{d\sigma_{{\gamma}p\to {\eta^\prime}p}(E_{\gamma})}
{dE_{\rm kin}^{\eta^\prime}d{\bf \Omega}_{N}}\right>_{{\Delta}{\bf \Omega}_N}=
\frac{{\pi}\theta(1-|x_0|)\theta(\omega-E_{\rm kin}^{\rm N,G})}
{(\cos{1}^{\circ}-\cos{11}^{\circ})I_2(s,m_N,m_{\eta^\prime})E_{\gamma}}
\end{equation}
$$
\times
{\frac{d\sigma_{{\gamma}p\rightarrow {\eta^\prime}p}[{\sqrt{s}},m_{{\eta^\prime}},
m_{N},{\theta_{\eta^\prime}^*}(x_0)]}
{d{\bf \Omega}_{\eta^\prime}^{*}}}
\theta(\cos1^{\circ}-\cos{\theta}_{\bf Q})
\theta(\cos{\theta}_{\bf Q}-\cos11^{\circ}),
$$
where
\begin{equation}
x_0=\frac{(p_{\gamma}^2+p_{\eta^\prime}^{2}-{\omega}^2-2m_N{\omega})}{(2p_{\gamma}p_{\eta^\prime})},\\\
x=\cos{\theta_{\eta^\prime}},\\\
\cos{\theta_{\bf Q}}=\frac{p_{\gamma}-p_{\eta^\prime}x_0}{\sqrt{{\omega}^2+2m_N{\omega}}},\\\
s=(E_{\gamma}+m_N)^2-p_{\gamma}^2.
\end{equation}

Let us consider now the two-step rescattering mechanism for the $(\gamma,{\eta^\prime}p)$
reaction on nuclei.

\section*{2.2. Two-step ${\eta^\prime}p$ production mechanism}

\hspace{1.5cm} The kinematical arguments led to the suggestion that the following two-step production
processes with a nucleon in an intermediate states may contribute to the ${\eta^\prime}p$ pair
production in ${\gamma}A$ interactions at bombarding energies of interest. An incident photon produces
in the first inelastic collision with an intranuclear nucleon an $\eta^\prime$ meson and nucleon via the
elementary reactions:
\begin{equation}
\gamma+N \to \eta^\prime +N,\\\\\ N=p,n.
\end{equation}
Then, the intermediate nucleon scatters elastically on another target nucleon (neutron if $N=p$,
or proton if $N=n$)
\begin{equation}
p+n \to p +n,
\end{equation}
\begin{equation}
n+p \to p +n
\end{equation}
in such a way that the final proton turns out to be in the Mini-TAPS acceptance window.
It should be noted that we neglected in our consideration the $pp \to pp$ rescatterings due to the
specific event selection rules of the present experiment, which exclude events with two charged
hits in the full solid angle.

  Accounting for the medium effects on the $\eta^\prime$ mass on the same footing as that employed
in calculating the ${\eta^\prime}p$ production cross section (10) from the primary process (1)
and assuming for the sake of numerical simplicity that the comparatively slow nucleons participating
in the elementary processes (31), (32) are in a potential well, obtained in the noninteracting Fermi-gas
model, of depth $U_N=-({\bar E}_F+\epsilon_A)=-31$ MeV at total binding energy per nucleon
$\epsilon_A=7$ MeV as well as
using the results given in [19], we get the following expression for the ${\eta^\prime}p$ production
exclusive differential cross section from the creation/rescattering sequence (30)--(32):
\begin{equation}
\frac{d\sigma_{{\gamma}A\to {\eta^\prime}pX}^{({\rm sec})}
(E_{\gamma})}
{d{\bf p}_{\eta^\prime}d{\bf p}_{N}}=I_{V}^{({\rm sec})}[A]
\left(\frac{ZN}{A^2}\right)\int d{\bf p}_{N}^{'}
[\left<\frac{d\sigma_{{\gamma}p\to {\eta^\prime}p}({\bf p}_{\gamma};
{\bf p}_{\eta^\prime}^{'},{\bf p}_N^{'})}{d{\bf p}^{\prime}_{\eta^\prime}d{\bf p}_N^{'}}\right>_A
\left<\frac{d\sigma_{pn\to pn}(p,{\bf p}_{N}^{'}\to p,{\bf p}_N^{f})}{d{\bf p}_N^{f}}\right>_A
\end{equation}
$$
+
\left<\frac{d\sigma_{{\gamma}n\to {\eta^\prime}n}({\bf p}_{\gamma};
{\bf p}_{\eta^\prime}^{'},{\bf p}_N^{'})}{d{\bf p}^{\prime}_{\eta^\prime}d{\bf p}_N^{'}}\right>_A
\left<\frac{d\sigma_{np\to pn}(n,{\bf p}_{N}^{'}\to p,{\bf p}_N^{f})}{d{\bf p}_N^{f}}\right>_A]
\frac{d{\bf p}_{\eta^\prime}^{'}}{d{\bf p}_{{\eta^\prime}}}
\frac{d{\bf p}_{N}^{f}}{d{\bf p}_{N}},
$$
where
\footnote{$^)$It should be pointed out that due to the moderate dependence of the quantity
$I_{V}^{({\rm sec})}[A]$ on the $\eta^\prime$, intermediate nucleon and detected proton
production angles in the lab frame, we assume, writing
the formula (34), that they move in the nucleus in forward direction.
Evidently, this is a good approximation for the latter one due to the kinematics of the
experiment under consideration. Also, that is well justified for the intermediate nucleon
and $\eta^\prime$ meson, since in line with the kinematic calculations and the results, given in [28],
their laboratory creation angles in ${\gamma}N \to {\eta^\prime}N$ reaction proceeding on the target
nucleon being at rest are $\le$ 35$^{\circ}$ at beam energies $\le$ 2.0 GeV giving the main contribution
to the ${\eta^\prime}p$ production cross section in the incident photon energy range of our interest.}$^)$
\begin{equation}
I_{V}^{({\rm sec})}[A]=2{\pi}A^2\int\limits_{0}^{R}r_{\bot}dr_{\bot}
\int\limits_{-\sqrt{R^2-r_{\bot}^2}}^{\sqrt{R^2-r_{\bot}^2}}dz
\rho(\sqrt{r_{\bot}^2+z^2})
\int\limits_{0}^{\sqrt{R^2-r_{\bot}^2}-z}dl
\rho(\sqrt{r_{\bot}^2+(z+l)^2})
\end{equation}
$$
\times
\exp{\left[-\sigma_{NN}^{\rm tot}A\int\limits_{z}^{z+l}
\rho(\sqrt{r_{\bot}^2+x^2})dx
-\sigma_{pN}^{\rm tot}A\int\limits_{z+l}^{\sqrt{R^2-r_{\bot}^2}}
\rho(\sqrt{r_{\bot}^2+x^2})dx\right]}
$$
$$
\times
\exp{\left[-\sigma_{{\eta^\prime}N}A\int\limits_{z}^{\sqrt{R^2-r_{\bot}^2}}
\rho(\sqrt{r_{\bot}^2+x^2})dx\right]}.
$$
Here, $\sigma_{NN}^{\rm tot}$ and $\sigma_{pN}^{\rm tot}$
are the effective total cross sections of the $NN$ and $pN$ interactions
\footnote{$^)$Taking into account that these cross sections depend weakly on the collision energy [29],
we use in the subsequent calculations the value of $\sigma_{NN}^{\rm tot}=40$ mb and
$\sigma_{pN}^{\rm tot}=40$ mb for all intermediate nucleon and final proton momenta involved.}$^)$
.
The quantity
$\left<d\sigma_{{\gamma}p\to {\eta^\prime}p}({\bf p}_{\gamma};
{\bf p}_{\eta^\prime}^{'},{\bf p}_N^{'})/d{\bf p}^{\prime}_{\eta^\prime}d{\bf p}_N^{'}\right>_A$,
entering into equation (33), is defined before by the formulas (13), (14) in which, according to the
above mentioned, one has to put:
\begin{equation}
<m^*_{N}({\bf p}_N^{'2})>=m_{N}+\frac{\sqrt{m_N^2+{{\bf p}_N^{'2}}}}{m_N}U_N.
\end{equation}
An analogous definition can be given for the differential cross section
$\left<d\sigma_{{\gamma}n\to {\eta^\prime}n}({\bf p}_{\gamma};
{\bf p}_{\eta^\prime}^{'},{\bf p}_N^{'})/d{\bf p}^{\prime}_{\eta^\prime}d{\bf p}_N^{'}\right>_A$.

 Taking into consideration that the energy conservation in the in-medium elastic elementary processes
(31), (32) looks like that for the free particles due to the cancelation of nuclear potentials $U_N$,
we assume that the nucleons participating in these processes are on-mass shell. Then, for example,
the averaged differential cross section
$\left<d\sigma_{pn\to pn}(p,{\bf p}_{N}^{'}\to p,{\bf p}_N^{f})/d{\bf p}_N^{f}\right>_A$, entering into
the equation (33), can be written in the following form [30]:
\begin{equation}
\left<\frac{d\sigma_{pn\to pn}(p,{\bf p}_{N}^{'}\to p,{\bf p}_N^{f})}{d{\bf p}_N^{f}}\right>_A=
\frac{3}{v_N^{'}4{\pi}{\bar p}_F^3}\int \int v_{\rm rel}
\frac{d\sigma_{pn\to pn}(p,{\bf p}_{N}^{'};n,{\bf p}^{''}\to p,{\bf p}_N^{f};n,{\bf p}^{'''})}
{d{\bf p}_N^{f}d{\bf p}^{'''}}
\end{equation}
$$
\times
\theta({\bar p}_F-p^{''})\theta(p^{'''}-{\bar p}_F)d{\bf p}^{''}d{\bf p}^{'''};
$$
where
\begin{equation}
\frac{d\sigma_{pn\to pn}(p,{\bf p}_{N}^{'};n,{\bf p}^{''}\to p,{\bf p}_N^{f};n,{\bf p}^{'''})}
{d{\bf p}_N^{f}d{\bf p}^{'''}}=
\left(\frac{\tilde s}{v_{\rm rel}}\right)
\frac{1}{\sqrt{m_N^2+{\bf p}_N^{'2}}\sqrt{m_N^2+{\bf p}^{''2}}\sqrt{m_N^2+{\bf p}_N^{f2}}
\sqrt{m_N^2+{\bf p}^{'''2}}}
\end{equation}
$$
\times
\frac{d\sigma_{pn \to pn}(|{\bf p}_N^{'}-{\bf p}^{''}|,\theta_{\rm cms})}
{d{\bf \Omega}_{\rm cms}}
$$
$$
\times
 \delta(\sqrt{m_N^2+{\bf p}_N^{'2}}+\sqrt{m_N^2+{\bf p}^{''2}}-\sqrt{m_N^2+{\bf p}_N^{f2}}
-\sqrt{m_N^2+{\bf p}^{'''2}})\delta({\bf p}_N^{'}+{\bf p}^{''}-{\bf p}_{N}^{f}-{\bf p}^{'''});
$$
\begin{equation}
v_N^{'}=|{\bf p}_N^{'}|/\sqrt{m_N^2+{\bf p}_N^{'2}},\\\
v_{\rm rel}=I/(\sqrt{m_N^2+{\bf p}_N^{'2}}\sqrt{m_N^2+{\bf p}^{''2}}),\\\
I=\sqrt{({\hat p}_N^{'}{\hat p}^{''})^2-m_N^4},
\end{equation}
\begin{equation}
{\tilde s}=({\hat p}_N^{'}+{\hat p}^{''})^2,\\\
{\hat p}_N^{'}=(\sqrt{m_N^2+{\bf p}_N^{'2}},{\bf p}_N^{'}),\\\
{\hat p}^{''}=(\sqrt{m_N^2+{\bf p}^{''2}},{\bf p}^{''}).
\end{equation}
Here, $d\sigma_{pn \to pn}/d{\bf \Omega}_{\rm cms}$ is the differential cross section of the
elastic rescattering process $pn \to pn$ in the $pn$ c.m.s.,
and $|{\bf p}_{N}^{f}|>{\bar p}_F$, $|{\bf p}_{N}^{'}|>{\bar p}_F$.
The quantity $\left<d\sigma_{np\to pn}(n,{\bf p}_{N}^{'}\to p,{\bf p}_N^{f})/d{\bf p}_N^{f}\right>_A$ in (33)
is described by the same formulas as those of (36)--(39) given above, but in which the scattering angle
$\theta_{\rm cms}$ should be replaced by $\pi-\theta_{\rm cms}$.

  Let us now simplify the expression (36) describing the differential cross section of
elastic  $pn \to pn$ scattering averaged over Fermi motion.
Considering that $|{\bf p}_{N}^{'}| \gg {\bar p}_F$, we can recast this expression into
the form [30]:
\begin{equation}
\left<\frac{d\sigma_{pn\to pn}(p,{\bf p}_{N}^{'}\to p,{\bf p}_N^{f})}{d{\bf p}_N^{f}}\right>_A=
\frac{d\sigma_{pn \to pn}(|{\bf p}_N^{'}|,\theta_{\rm cms})}
{d{\bf \Omega}_{\rm cms}}
K_q(E_{\rm kin}^{'} \to E_{\rm kin}^{f},{\bf \Omega}_{N}^{'} \to {\bf \Omega}_{N}^{f})
\frac{dE_{\rm kin}^{f}d{\bf \Omega}_{N}^{f}}{d{\bf p}_N^{f}},
\end{equation}
where
\begin{equation}
E_{\rm kin}^{'}=\sqrt{m_N^2+{\bf p}_N^{'2}}-m_N,\\\
E_{\rm kin}^{f}=\sqrt{m_N^2+{\bf p}_N^{f2}}-m_N,\\\
{\bf \Omega}_{N}^{'}={\bf p}_N^{'}/|{\bf p}_N^{'}|,\\\ {\bf \Omega}_{N}^{f}={\bf p}_N^{f}/|{\bf p}_N^{f}|
\end{equation}
and indicatrix of scattering $K_q(E^{'} \to E,{\bf \Omega}^{'} \to {\bf \Omega})$ is defined as follows [30]:
\begin{equation}
K_q(E^{'} \to E,{\bf \Omega}^{'} \to {\bf \Omega})=
\frac{3\theta(1-2{\tilde z})}{2E^{'}{\tilde \omega}\sqrt{{\bar E}_F}}\theta(\kappa)
\theta\left(\frac{E^{'}-E}{{\bar E}_F}\right)
\left[\kappa\theta\left(\frac{E^{'}-E}{{\bar E}_F}-\kappa\right)+
\frac{E^{'}-E}{{\bar E}_F}\theta\left(\kappa-\frac{E^{'}-E}{{\bar E}_F}\right)\right];
\end{equation}
\begin{equation}
{\tilde \omega}=\left|\frac{{\bf \Omega}}{\sqrt{\nu}}-\frac{{\bf \Omega}^{'}}{\sqrt{\nu^{'}}} \right|,\\\
\nu=E\left(1+\frac{E}{2m_N}\right),
\end{equation}
\begin{equation}
\kappa=1-\frac{2}{{\tilde \omega}^2{\bar E}_F(1-{\tilde z}+\sqrt{1-2{\tilde z}})}
\left({\bf \Omega}^{'}{\bf \Omega}-\frac{E}{E^{'}}\sqrt{\frac{\nu^{'}}{\nu}}\right)^2,
\end{equation}
\begin{equation}
{\tilde z}=\frac{E^{'}-E}{m_N}\frac{1}{{\tilde \omega}^2\sqrt{\nu\nu^{'}}}
\left({\bf \Omega}^{'}{\bf \Omega}-\frac{E}{E^{'}}\sqrt{\frac{\nu^{'}}{\nu}}\right).
\end{equation}
It should be pointed out that the kinetic energies $E_{\rm kin}^{'}$ and $E_{\rm kin}^{f}$ in (40) are counted
from the bottom of the nuclear potential well. Therefore, the kinetic energy $E_{\rm kin}^{f}$ of the detected
proton inside the nuclear matter is related to the vacuum one $E_{\rm kin}^{N}$ by the expression:
\begin{equation}
E_{\rm kin}^{f}+U_N=E_{\rm kin}^{N}.
\end{equation}

 Substituting the equations (40), (46) into (33) and neglecting the anisotropy of the cross section
$d\sigma_{pn \to pn}/d{\bf \Omega}_{\rm cms}$ and any difference in the directions of the in-medium
and vacuum final proton solid angles ${\bf \Omega}_{N}^{f}$ and ${\bf \Omega}_{N}$ as well as
accounting for that the indicatrix of scattering $K_q(E^{'} \to E,{\bf \Omega}^{'} \to {\bf \Omega})$
is a function of the scalar product ${\bf \Omega}^{'}{\bf \Omega}$ and performing the respective integrations,
we can get the following expression for the differential cross section for photoproduction of ${\eta^\prime}$
mesons off nuclei in coincidence with protons from the production/rescattering chain (30)--(32) in the kinematics
of the CBELSA/TAPS experiment of our interest:
\begin{equation}
\left<\frac{d\sigma_{{\gamma}A\to {\eta^\prime}pX}^{({\rm sec})}(E_{\gamma})}
{dE_{\rm kin}^{\eta^\prime}d{\bf \Omega}_{N}}\right>_{{\Delta}{\bf \Omega}_N}=
\left(\frac{1}{{\Delta}{\bf \Omega}_N}\right)\left(\frac{ZN}{A^2}\right)
I_{V}^{({\rm sec})}[A]\sum_{N=p,n}
\int\limits_{-1}^{1}d\cos{{\theta_{\eta^\prime}}}\int\limits_{0}^{2\pi}d{\phi}_{\eta^\prime}
\end{equation}
$$
\times
\left(\frac{p_{\eta^\prime}^{'}}{4Q}\right)\int\limits_{0}^{p_t^{\rm max}}
\int\limits_{0}^{2\pi}\int\limits_{E_{\rm min}}^{E_{\rm max}}P_A(p_t,E)p_tdp_td\phi_tdE
$$
$$
\times
\frac{\theta(1-|x_0|)\theta(\omega+E_t)\sigma_{pn \to pn}(\sqrt{Q^2+p_t^2+2Qp_tx_0})}
{I_2[s(x_0,p_t,\phi_t,E),<m^*_{N}(Q^2+p_t^2+2Qp_tx_0)>,<m^*_{{\eta^\prime}}>]
F_{\rm DIF}^{({\rm sec})}(x_0,p_t,Q)}
$$
$$
\times
{\frac{d\sigma_{{\gamma}N \to {\eta^\prime}N}[{\sqrt{s(x_0,p_t,\phi_t,E)}},<m^*_{{\eta^\prime}}>,
<m^*_{N}(Q^2+p_t^2+2Qp_tx_0)>,{\theta_{\eta^\prime}^*}]}
{d{\bf \Omega}_{\eta^\prime}^{*}}}
$$
$$
\times
\int\limits_{E_{\rm kin}^{\rm N,G}-U_N}^{\sqrt{m_N^2+Q^2+p_t^2+2Qp_tx_0}-m_N}dE_{\rm kin}^{f}
\int\limits_{{\Delta}{\bf \Omega}_N}d{\bf \Omega}_{N}^{f}
\theta\left[\sqrt{m_N^2+Q^2+p_t^2+2Qp_tx_0}-m_N-(E_{\rm kin}^{\rm N,G}-U_N)\right]
$$
$$
\times
K_q\left(\sqrt{m_N^2+Q^2+p_t^2+2Qp_tx_0}-m_N \to E_{\rm kin}^{f},
\frac{{\bf Q}+{\bf p}_t}{|{\bf Q}+{\bf p}_t|} \to {\bf \Omega}_{N}^{f}\right),
$$
where
\begin{equation}
F_{\rm DIF}^{({\rm sec})}(x_0,p_t,Q)=\left|1+\left(m_{N}+\frac{\sqrt{m_N^2+Q^2+p_t^2+2Qp_tx_0}}{m_N}U_N\right)
\frac{U_N}{m_N\sqrt{m_N^2+Q^2+p_t^2+2Qp_tx_0}}\right|,
\end{equation}
\begin{equation}
x_0=\frac{t_0^2-m_N^2-Q^2-p_t^2}{2Qp_t},\\\\\
t_0=\frac{\sqrt{U_N^2+\left(1+\frac{U_N^2}{m_N^2}\right)(\omega+E_t)^2}-U_N}
{\left(1+\frac{U_N^2}{m_N^2}\right)};
\end{equation}
\begin{equation}
d{\bf \Omega}_{N}^{f}=\sin{\theta_N^f}d\theta_N^fd\phi_N^f,
\end{equation}
\begin{equation}
\frac{({\bf Q}+{\bf p}_t){\bf \Omega}_{N}^{f}}{|{\bf Q}+{\bf p}_t|}=
\frac{({\bf p}_{\gamma}-{\bf p}_{\eta^\prime}^{'}+{\bf p}_t){\bf \Omega}_{N}^{f}}{|{\bf Q}+{\bf p}_t|}=
\frac{p_{\gamma}\cos{\theta_N^f}-p_{\eta^\prime}^{'}\cos{\theta_{{\bf p}_{\eta^\prime}^{'}{\bf \Omega}_{N}^{f}}}+
p_t\cos{\theta_{{\bf p}_t{\bf \Omega}_N^f}}}
{\sqrt{Q^2+p_t^2+2Qp_tx_0}},
\end{equation}
\begin{equation}
\cos{\theta_{{\bf p}_{\eta^\prime}^{'}{\bf \Omega}_{N}^{f}}}=
\cos{\theta_{{\eta^\prime}}}\cos{\theta_N^f}+\sin{\theta_{{\eta^\prime}}}\sin{\theta_N^f}
\cos{(\phi_{{\eta^\prime}}-\phi_N^f)},
\end{equation}
\begin{equation}
\cos{\theta_{{\bf p}_t{\bf \Omega}_N^f}}=
\cos{\theta_{x}}\cos{\theta_N^f}+\sin{\theta_{x}}\sin{\theta_N^f}
\cos{(\phi_N^f-\phi_t^{'})}.
\end{equation}
Here, $\phi_N^f$ and $\phi_t^{'}$ are the azimuthal angles of ${\bf \Omega}_{N}^{f}$ and ${\bf p}_t$
in the reference frame with the $z$ axis directed along the photon momentum ${\bf p}_{\gamma}$ and
$\sigma_{pn \to pn}(|{\bf p}_N^{'}={\bf Q}+{\bf p}_t|)$ is the free total cross section of the elastic
$pn \to pn$ rescattering process. In our calculations we adopted for this cross section the following fit
of the available data:
\begin{equation}
\sigma_{pn \to pn}(|{\bf p}_N^{'}|)=27.15+(1.8/E_{\rm kin}^{'})+(0.24/E_{\rm kin}^{'2})~[\rm mb] \\\
for~0.04~{\rm GeV} \le E_{\rm kin}^{'} \le 1.0~{\rm GeV},
\end{equation}
where the nucleon kinetic energy $E_{\rm kin}^{'}$ ($E_{\rm kin}^{'}=\sqrt{m_N^2+{\bf p}_N^{'2}}-m_N$)
is measured in GeV.
\begin{figure}[htb]
\begin{center}
\includegraphics[width=12.0cm]{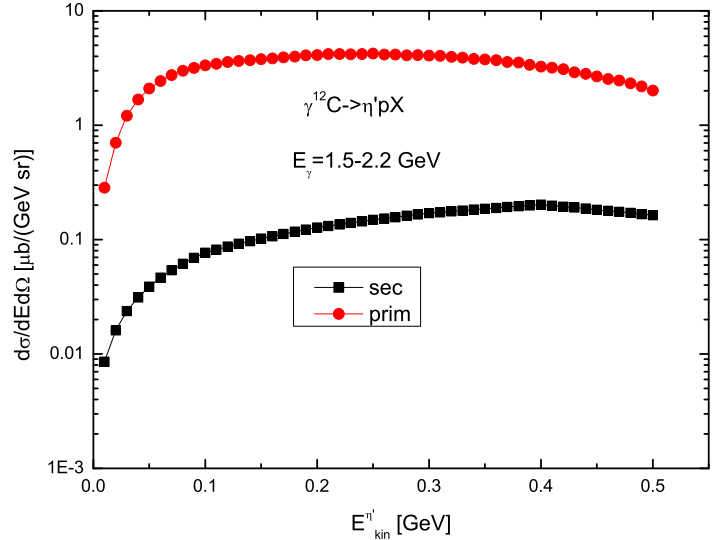}
\vspace*{-2mm} \caption{(color online) Differential cross sections as functions of the
$\eta^\prime$ vacuum kinetic energy for photoproduction
of ${\eta^\prime}$ mesons in the full solid angle off the carbon target nucleus from primary (1)
and secondary (30)--(32) processes in coincidence with protons, going into the laboratory polar angular range of $1^{\circ}$--$11^{\circ}$, after averaging over the incident photon energy range of 1.5--2.2 GeV
in the scenario without $\eta^\prime$ meson mass shift.}
\label{void}
\end{center}
\end{figure}

\section*{3. Results and discussion}

\hspace{1.5cm} At first, we consider the ${\eta^\prime}p$ production cross sections from the one-step
and two-step ${\eta^\prime}p$ creation mechanisms in ${\gamma}$$^{12}$C collisions, calculated on the
basis of equations (19) and (47) for photon energies of 1.5--2.2 GeV. These cross sections are shown
in figure 1. One can see that the primary ${\gamma}p \to {\eta^\prime}p$ channel plays the dominant role
for the kinematical conditions of the CBELSA/TAPS experiment. This gives confidence to us that the secondary
processes (30)--(32) can be ignored in our subsequent calculations.
\begin{figure}[!h]
\begin{center}
\includegraphics[width=12.0cm]{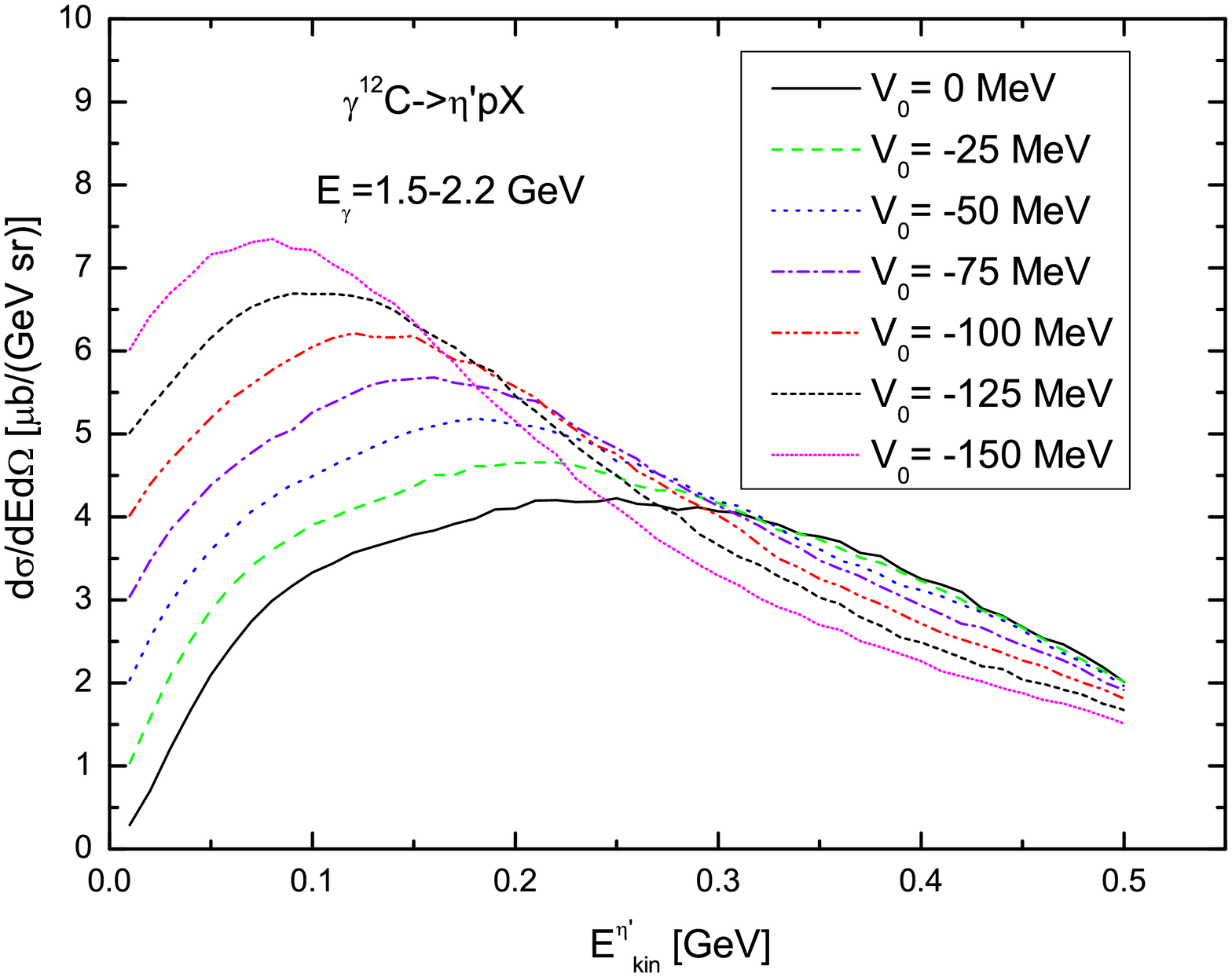}
\vspace*{-2mm} \caption{(color online) Differential cross section as a function of the
$\eta^\prime$ vacuum kinetic energy for
photoproduction of $\eta^\prime$ mesons in the full solid angle off the carbon target nucleus
from primary channel (1) in coincidence with protons, going into the laboratory polar angular range of $1^{\circ}$--$11^{\circ}$ and feeling inside the nucleus a momentum-dependent
nuclear potential (8), after averaging over the incident photon energy range of 1.5--2.2 GeV
for different in-medium $\eta^\prime$ mass shifts at normal nuclear density indicated in the inset.}
\label{void}
\end{center}
\end{figure}
\begin{figure}[!h]
\begin{center}
\includegraphics[width=12.0cm]{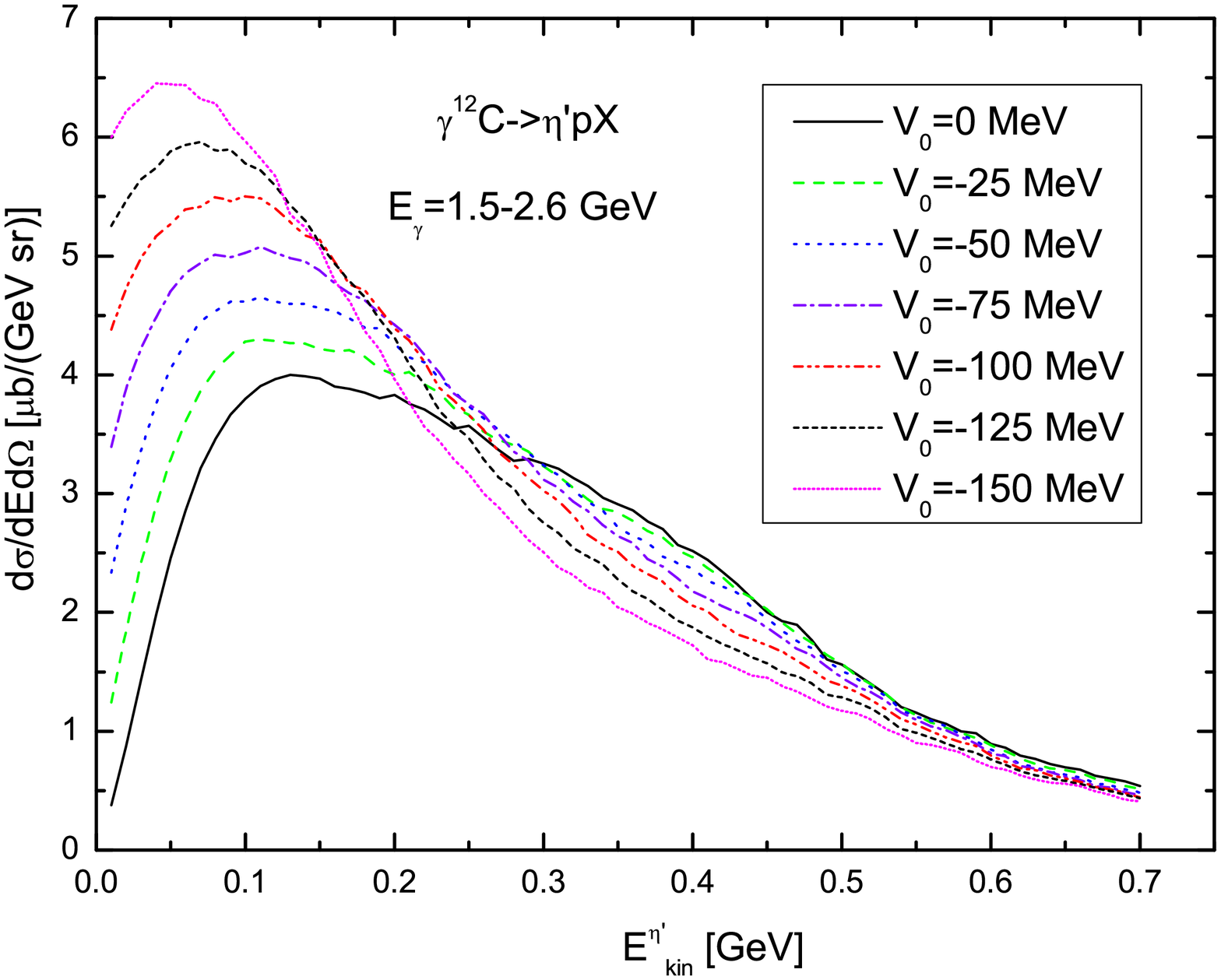}
\vspace*{-2mm} \caption{(color online) The same as in figure 2, but for the interaction of
photons of energies of 1.5--2.6 GeV with the carbon target nucleus.}
\label{void}
\end{center}
\end{figure}
\begin{figure}[!h]
\begin{center}
\includegraphics[width=12.0cm]{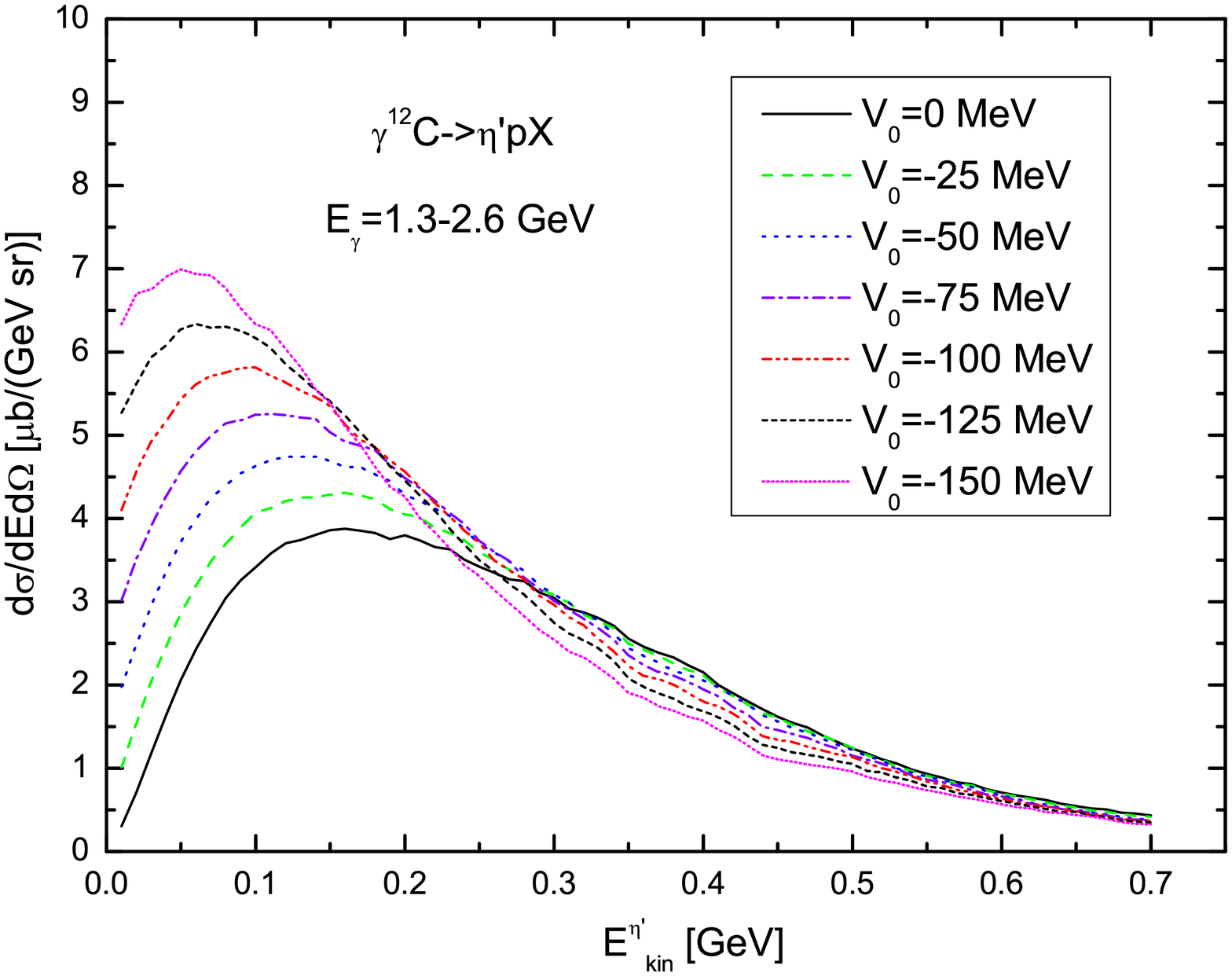}
\vspace*{-2mm} \caption{(color online) The same as in figure 2, but for the interaction of
photons of energies of 1.3--2.6 GeV with the carbon target nucleus.}
\label{void}
\end{center}
\end{figure}
\begin{figure}[!h]
\begin{center}
\includegraphics[width=12.0cm]{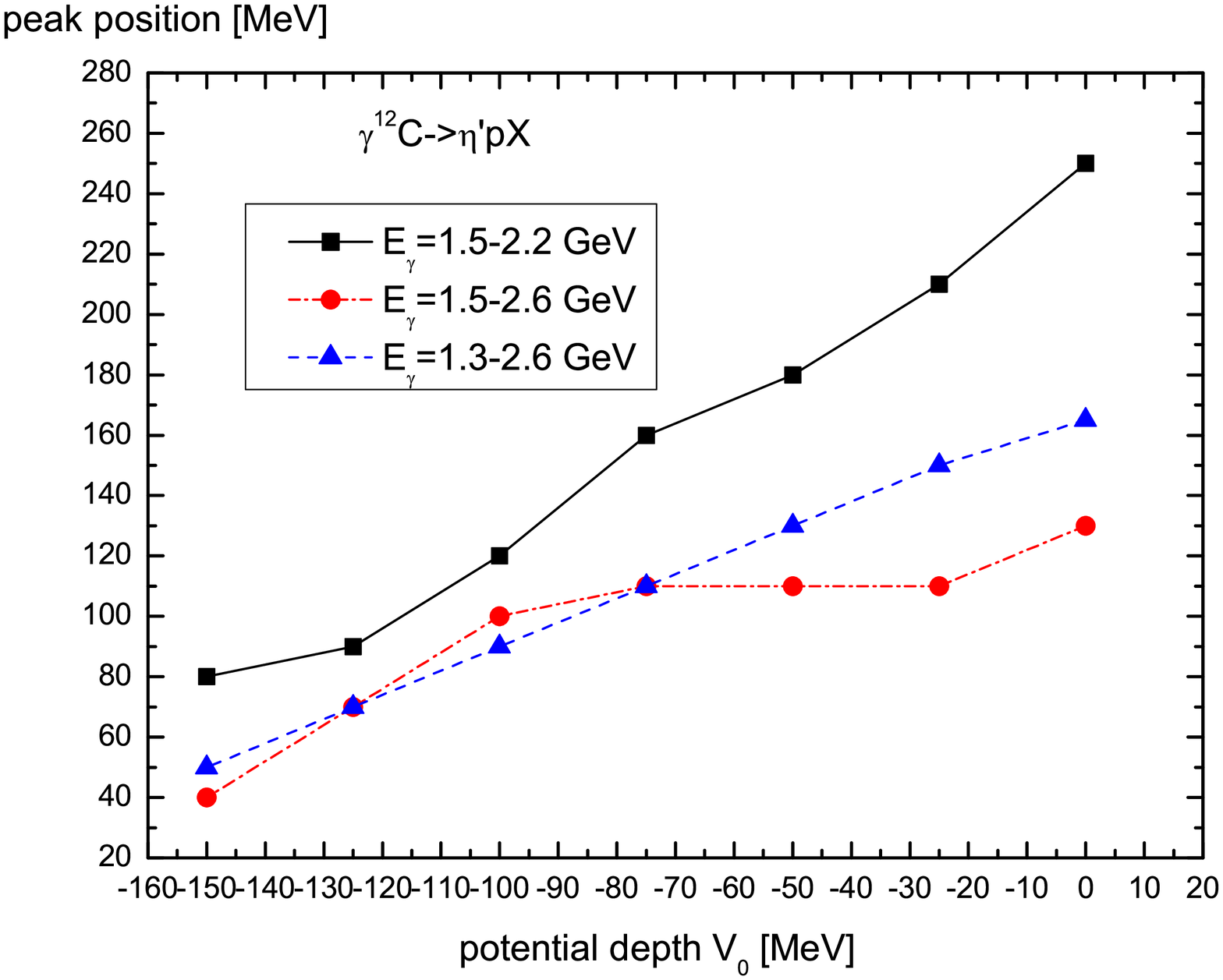}
\vspace*{-2mm} \caption{(color online) Peak positions in the $\eta^\prime$ meson kinetic energy
distributions shown in figures 2, 3 and 4, as functions of the in-medium mass shift at normal
nuclear density. The lines are to guide the eyes.}
\label{void}
\end{center}
\end{figure}

    Now, we concentrate on the differential cross sections for photoproduction of $\eta^\prime$ mesons off
$^{12}$C from the primary process (1) in coincidence with protons, going into the polar angular range of
$1^{\circ}$--$11^{\circ}$ covered by the Mini-TAPS forward array. They were calculated on the basis of
equation (19) for three initial photon energy intervals of 1.5--2.2 GeV, 1.5--2.6 GeV and 1.3--2.6 GeV
as well as for seven adopted scenarios for the $\eta^\prime$ in-medium mass shift, and are given in
figures 2, 3 and 4, respectively. It can be seen that there are clear correlations between the peaks
in the $\eta^\prime$ kinetic energy distributions, shown in these figures, and the $\eta^\prime$
in-medium mass shift, namely: the peaks move to lower energies with decreasing this shift down to -150 MeV.
To illustrate these findings more quantitatively,
these correlations are plotted in figure 5. It is helpful to point out that they have different slopes and
that the incident photon energy range of 1.5--2.6 GeV is not optimal for determining the $\eta^\prime$
in-medium mass shift. The highest sensitivity to this shift we have for the photon energy range of
1.5--2.2 GeV, but because of the expected higher statistics the largest bin of 1.3--2.6 GeV is preferred.
The reason for the different slopes in the correlations shown in figure 5 is the following. Let us start
with the initial photon energy interval of 1.5--2.2 GeV. If we increase upper energy limit to 2.6 GeV,
maintaining the lower one of 1.5 GeV, we add, as calculations showed, events mainly with low $\eta^\prime$
kinetic energies from the kinematic branch with $\eta^\prime$ mesons going backward and protons going
forward in cm system. This shifts the respective maxima towards the lower kinetic energy values. If we then
keep the upper photon energy limit of 2.6 GeV and decrease the lower one to 1.3 GeV, then we add mainly
events from the region where the above kinematic branch and that corresponding to $\eta^\prime$ mesons
going forward and protons going backward in cm system merge. This shifts the maxima again to higher
kinetic energy values. Therefore, a comparison above results with the
experimentally determined peak position in the $\eta^\prime$ kinetic energy distribution under consideration
will allow one to determine a possible $\eta^\prime$ meson mass shift in cold nuclear matter. It should be
noted that an analogous possibility was recently realized for the $\omega$ mesons in [31].
\begin{figure}[!h]
\begin{center}
\includegraphics[width=12.0cm]{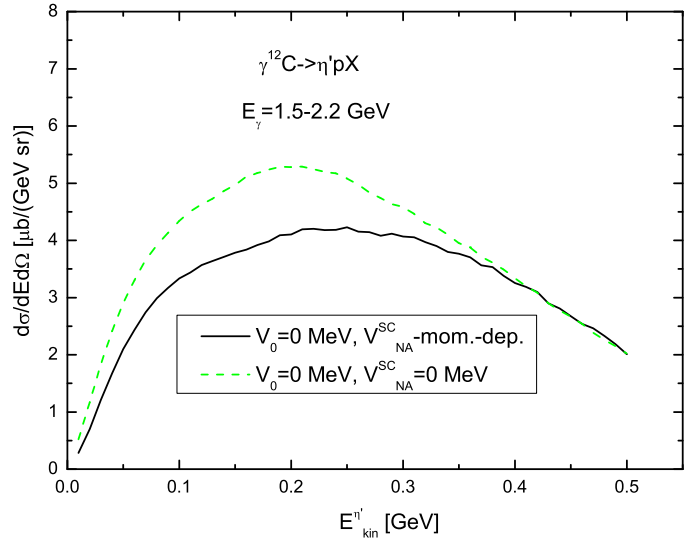}
\vspace*{-2mm} \caption{(color online) Differential cross section as a function of the
$\eta^\prime$ vacuum kinetic energy for
photoproduction of $\eta^\prime$ mesons in the full solid angle off the carbon target nucleus
from primary channel (1) in coincidence with protons, going into the laboratory polar angular range of
$1^{\circ}$--$11^{\circ}$ as well as feeling and not seeing a momentum-dependent
nuclear potential (8) inside the nucleus (solid and dashed lines), respectively,
after averaging over the incident photon energy range of 1.5--2.2 GeV in the scenario without
$\eta^\prime$ in-medium mass shift.}
\label{void}
\end{center}
\end{figure}

     In addition, we have investigated the influence of the scalar momentum-dependent potential (8),
which the outgoing participant proton sees inside the carbon nucleus, on the ${\eta^\prime}p$ yield.
Figure 6 shows that the inclusion of this potential results in a reduction of the $\eta^\prime$ kinetic
energy distribution on carbon by a factor of about 1.3 at low $\eta^\prime$ kinetic energies
as well as in a shift of the peak of this distribution by about of 50 MeV to higher energies. This is
due to the fact that in a ($\gamma$, ${\eta^\prime}p$) reaction on a nuclear target the low-energy
$\eta^\prime$ mesons are produced in the kinematics of our interest together with the high-energy
protons to balance the momentum of the incident photon beam (see the second footnote given above).
In view of equations (7) and (8), these
protons feel in the interior of the nucleus a repulsive potential, which leads to the above effects.
Therefore, one can conclude that to extract the reliable information on the $\eta^\prime$ in-medium
modification from the analysis of the measured $\eta^\prime$ kinetic energy distributions on nuclei,
like those just considered, it is important to account for the impact of the
momentum-dependent nuclear potential on the participant protons emerging from the nucleus.
\begin{figure}[!h]
\begin{center}
\includegraphics[width=12.0cm]{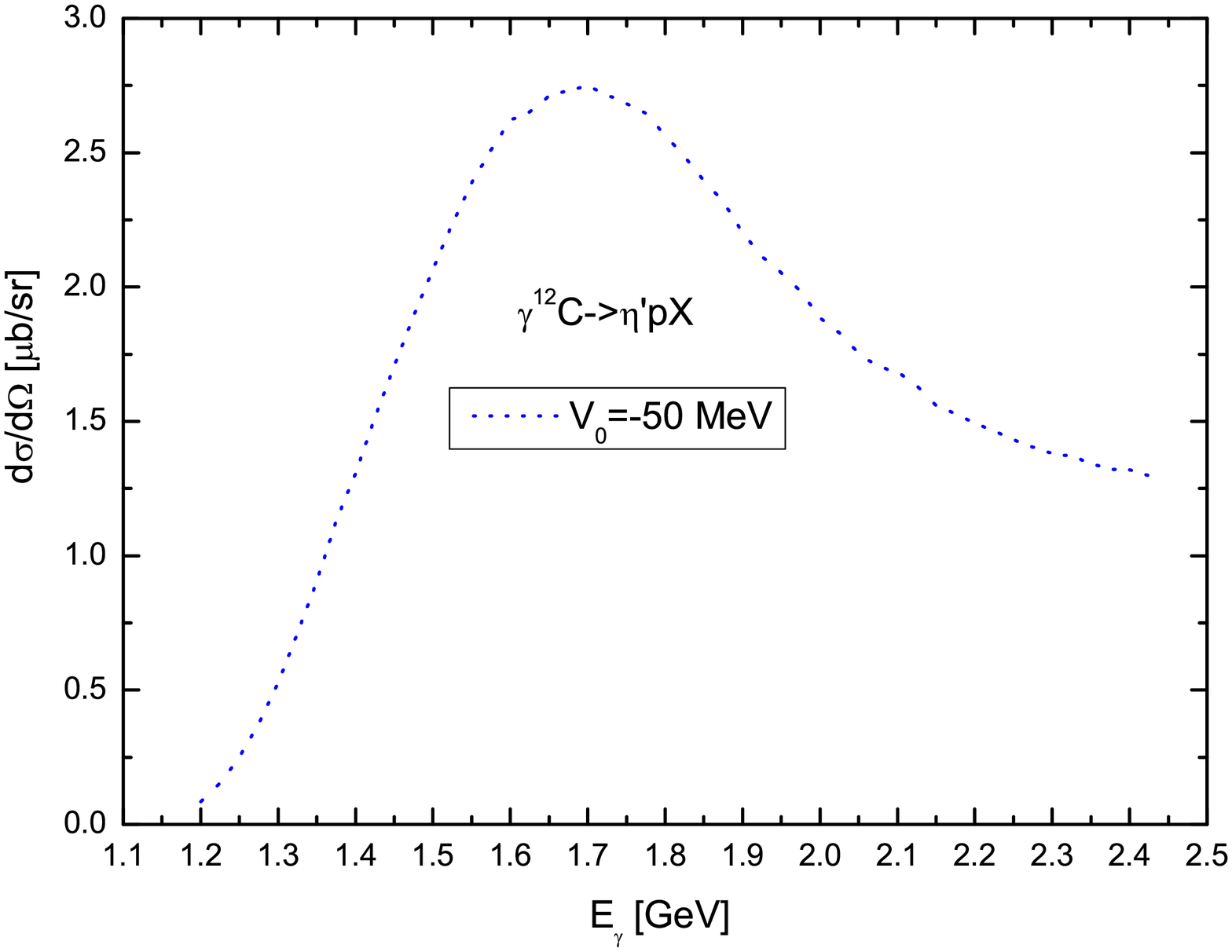}
\vspace*{-2mm} \caption{(color online) Excitation function for photoproduction of $\eta^\prime$
mesons in the full phase space off the carbon target nucleus from the primary channel (1)
in coincidence with protons, going into the laboratory polar angular range of $1^{\circ}$--$11^{\circ}$
and feeling inside the nucleus a momentum-dependent nuclear potential (8),
in the scenario with an $\eta^\prime$ in-medium mass shift at normal nuclear density of -50 MeV.}
\label{void}
\end{center}
\end{figure}
\begin{figure}[!h]
\begin{center}
\includegraphics[width=12.0cm]{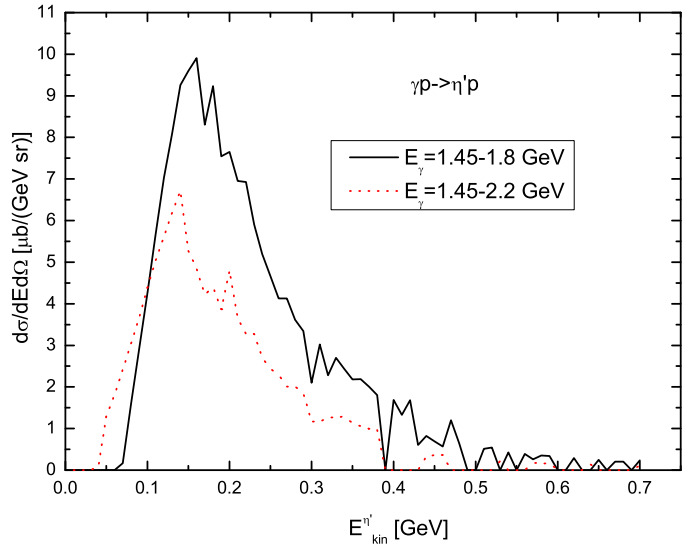}
\vspace*{-2mm} \caption{(color online) Differential cross section as a
function of the $\eta^\prime$ kinetic energy for photoproduction of $\eta^\prime$
mesons in the full solid angle off a free target proton at rest
in coincidence with protons, going into the laboratory polar angular range of $1^{\circ}$--$11^{\circ}$,
after averaging over the incident photon energy ranges of 1.45--1.8 GeV and 1.45--2.2 GeV
(solid and dotted lines, respectively).}
\label{void}
\end{center}
\end{figure}
\begin{figure}[!h]
\begin{center}
\includegraphics[width=12.0cm]{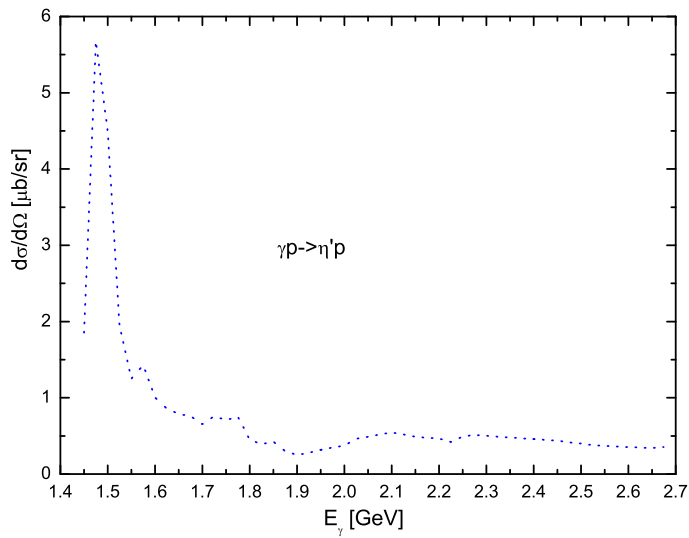}
\vspace*{-2mm} \caption{(color online) Excitation function for photoproduction of $\eta^\prime$
mesons in the full phase space off a free target proton at rest
in coincidence with protons, going into the laboratory polar angular range of $1^{\circ}$--$11^{\circ}$.}
\label{void}
\end{center}
\end{figure}

   In figure 7 we show the excitation function for photoproduction of $\eta^\prime$ mesons in
the full phase space off $^{12}$C from the primary channel (1) in coincidence with protons,
going into the laboratory polar angular range of $1^{\circ}$--$11^{\circ}$. It was calculated on
the basis of equation (19) in the scenario with $\eta^\prime$ in-medium mass shift $V_0=-50$ MeV,
which we believe is close to the reality [21]. It is seen that the calculation shows a peak at an
incident photon energy of 1.7 GeV. In view of the expected data from the CBELSA/TAPS experiment,
the results presented in figure 7 can be also used as an additional tool to those given before
for determining the possible $\eta^\prime$ in-medium mass shift.

   Finally, we consider the kinetic energy distribution and excitation function for $\eta^\prime$
mesons produced on a free target proton being at rest in reaction (1)
in coincidence with protons, going into the laboratory polar angular range of $1^{\circ}$--$11^{\circ}$.
They were calculated on the basis of equation (28), and are given in figures 8 and 9, respectively.
The importance of these predictions lies in the fact that their comparison with the corresponding
data from the CBELSA/TAPS experiment will provide a simple consistency check of the data analysis and our
calculations.

   Thus, we come to the conclusion that the considered above exclusive observables can be useful to help
determine the $\eta^\prime$ in-medium mass shift in cold nuclear matter.

\section*{4. Conclusions}

\hspace{1.5cm} With the aim of studying a possible shift of the $\eta^\prime$ meson mass in the
nuclear medium we have developed in this article an approach for the description of the
($\gamma$, ${\eta^\prime}p$) reaction on nuclei near the threshold in the kinematical conditions
of the Crystal Barrel/TAPS experiment, recently performed at ELSA. The approach accounts for both
a direct knock out process and a two-step mechanism where the knock out reaction
($\gamma$, ${\eta^\prime}N$) with $N=p, n$ is followed by an elastic rescattering of an intermediate
nucleon $N$ from another nucleon in the medium, leading to the emission of the detected proton, as
well as different scenarios of the $\eta^\prime$ in-medium mass shift. Calculations within
this approach have been performed for the $^{12}$C($\gamma$, ${\eta^\prime}p$) reaction for which
the respective data were collected by the CBELSA/TAPS Collaboration.
It was found
that the considered two-step mechanism plays a minor role in ${\eta^\prime}p$ photoproduction off a
carbon target in the chosen kinematics and, hence, it can be ignored here. The calculated exclusive
$\eta^\prime$ kinetic energy distributions from the primary ${\gamma}p \to {\eta^\prime}p$ channel show
a strong sensitivity to the $\eta^\prime$ in-medium mass shift in the studied ranges of the initial
photon energy. This provides the opportunity to determine it from a direct comparison of the results
of our calculations with the upcoming data from the respective CBELSA/TAPS experiment. It was also shown
that the above distributions are strongly sensitive as well to the momentum-dependent optical potential,
which the outgoing participant proton sees inside the carbon nucleus and, therefore, it should be accounted
for in the analysis of these data with the purpose to get information on the $\eta^\prime$ in-medium mass
shift in cold nuclear matter.
\\
\\
{\bf Acknowledgments}
\\
\\
The author is very grateful to V. Metag and M. Nanova for initiation of this study as well as
for fruitful discussions throughout it.
\\
\\

\end{document}